\documentclass[fleqn,10pt]{wlscirep}
\title{Generalised monogamy relation of convex-roof extended negativity in multi-level systems}

\author[1]{Tian Tian}
\author[2]{Yu Luo}
\author[3,*]{Yongming Li}

\affil[1,2,3]{College of Computer Science, Shaanxi Normal University, Xi'an, 710062, China}

\affil[*]{liyongm@snnu.edu.cn}

\begin{abstract}

In this paper, we investigate the generalised monogamy inequalities of convex-roof extended negativity (CREN) in multi-level systems. The generalised monogamy inequalities provide the upper and lower bounds of bipartite entanglement, which are obtained by using CREN and the CREN of assistance (CRENOA). Furthermore, we show that the CREN of multi-qubit pure states satisfies some monogamy relations. Additionally, we test the generalised monogamy inequalities for qudits by considering the partially coherent superposition of a
generalised W-class state in a vacuum and show that the generalised monogamy inequalities are satisfied in this case as well.
\end{abstract}
\begin{document}

\flushbottom
\maketitle
%
%
\thispagestyle{empty}

\section*{Introduction}

Quantum entanglement is one of the most important physical resources in quantum information processing~\cite{Horodecki09,Popescu14,Eisert10,Bennett93}.
As distinguished from classical correlations, quantum entanglement cannot be freely shared among many objects. We call this important phenomenon of quantum entanglement monogamy~\cite{Terhal04,Koashi04}. The property of monogamy may be as fundamental as the no-cloning theorem~\cite{Kay09}, which gives rise to structures of entanglement in multipartite settings~\cite{Coffman00,Osborne06}. Some monogamy inequalities have been studied to apply entanglement to more useful quantum information processing. The property of monogamy property has been considered in many areas of physics: it can be used to extract an estimate of the quantity of information about a secret key captured by an eavesdropper in quantum cryptography ~\cite{Bennett92,Barrett05}, as well as the frustration effects observed in condensed matter physics ~\cite{mar99,Ma11} and even black-hole physics~\cite{ka99,Lloyd14}.

The monogamy relation of entanglement is a way to characterise different types of entanglement distribution. The first monogamy relation was named the Coffman-Kundu-Wootters (CKW) inequality~\cite{Coffman00}. The monogamy property can be interpreted as the following statement: the amount of entanglement between $A$ and $B$ plus the amount of entanglement between $A$ and $C$ cannot be greater than the amount of entanglement between $A$ and the $BC$ pair. Osborne and Verstraete later proved that the CKW inequality also holds in an $n$-qubit system~\cite{Osborne06}. Other types of monogamy relations for entanglement were also proposed. Studies have found that the monogamy inequality holds in terms of some entanglement measures, negativity~\cite{Ou107}, squared CREN\cite{Kim09}, entanglement of formation~\cite{Oliveira14,Bai1401,L16}, $\rm{R\acute{e}nyi}$ entropy~\cite{Song16} and Tsallis entropy~\cite{Yu16,yu16}. 
The monogamy property of other physical resources, such as discord and steering~\cite{Bai13}, has also been discussed. There can be several inequivalent types of entanglement among the subsystems in multipartite quantum systems, and the amount of different types of entanglement might not be directly comparable to one another. Regula $et$ $al.$ studied multi-party quantum entanglement and found that there was strong monogamy~\cite{Choi15}. Additionally, generalised monogamy relations of concurrence for N-qubit systems were also proposed by Zhu $et$ $al$~\cite{Xue15}.

In this paper, we study the generalised monogamy inequalities of CREN in multi-qubit systems. We first recall some basic concepts of entanglement measures. Then, monogamy inequalities are given by the concurrence and negativity of the $n$-qubit entanglement. Furthermore, we consider some states in a higher-dimensional quantum system and find that the generalised monogamy inequalities also hold for these states. We specifically test the generalised monogamy inequalities for qudits by considering the partially coherent superposition of a generalised W-class state in a vacuum, and we show that the generalised monogamy inequalities are satisfied in this case as well. These relations also give rise to a type of trade-off in inequalities that is related to the upper and lower bounds of CRENOA. It shows the bipartite entanglement between $AB$ and the other qubits: especially under partition $AB$, a two-qubit system is different from the previous monogamy inequality that is typically used.

\section*{Results}

This paper is organised as follows: in the first subsection, we recall some basic concepts of concurrence and negativity. We present the monogamy relations of concurrence and negativity in the second subsection. In the third subsection, the generalised monogamy inequalities of CREN are given. The fourth subsection includes some examples that verify these results.

\section*{Preliminaries: concurrence and negativity }

For any bipartite pure state $|\psi\rangle_{AB}$ in a $d \otimes d' ~(d\leq d')$ quantum system with its Schmidt decomposition,
\begin{equation}
  |\psi\rangle_{AB}=\sum_{i=0}^{d-1}\sqrt{\lambda_i}|ii\rangle, ~~\lambda_i\geq0,\sum_{i=0}^{d-1}\lambda_i=1,
\end{equation}
the concurrence $\mathcal{C}(|\psi\rangle_{AB})$ is defined as~\cite{Wootters98}
\begin{equation}\label{eq:Corollary 1}
  \mathcal{C}(|\psi\rangle_{AB})=\sqrt{2[1-\rm{Tr}(\rho_A^2)]},
\end{equation}
where $\rho_A=tr_B(|\psi\rangle_{AB}\langle\psi|)$.
For any mixed state $\rho_{AB}$, its concurrence is defined as
\begin{equation}\label{eq:Corollary 2}
  \mathcal{C}(\rho_{AB})=\min\sum_{i}p_{i}\mathcal{C}(|\psi_i\rangle_{AB}),
\end{equation}
where the minimum is taken over all possible pure state
decompositions $\{p_{i},|\psi_i\rangle_{AB}\}$ of $\rho_{AB}$.

Similarly, the concurrence of assistance (COA) of $\rho_{AB}$ is defined as~\cite{Laustsen03}
\begin{equation}\label{eq:Corollary 3}
 \mathcal{C}_{a}(\rho_{AB})=\max\sum_{i}p_{i}\mathcal{C}(|\psi_i\rangle_{AB}),
\end{equation}
where the maximum is taken over all possible pure state
decompositions $\{p_{i},|\psi_i\rangle_{AB}\}$ of $\rho_{AB}$.

Another well-known quantification of bipartite entanglement is
negativity. For any bipartite pure state $|\psi\rangle_{AB}$, the negativity
$\mathcal{N}(|\psi\rangle_{AB})$ is
\begin{equation}\label{eq:Corollary 4}
  \mathcal{N}(|\psi\rangle_{AB})=2\sum_{i<j}\sqrt{\lambda_i\lambda_j}=(\rm{Tr}\sqrt{\rho_A})^2-1,
\end{equation}
where $\rho_A=tr_B(|\psi\rangle_{AB}\langle\psi|)$.

For any bipartite state $\rho_{AB}$ in the Hilbert space $\mathcal{H_A}\otimes\mathcal{H_B}$ negativity is defined as~\cite{Vidal02}
\begin{equation}
\mathcal{N}(\rho_{AB})=\frac{\|\rho^{T_A}_{AB}\|-1}{2},
\end{equation}
where $\rho^{T_A}_{AB}$ is a partial transposition with respect to the subsystem $A$, $\|X\|$ denotes the trace norm of $X$; i.e., $\|X\|\equiv \rm{Tr}\sqrt{XX^{\dag}}$. Negativity is a computable measure of entanglement, which is a convex function of $\rho_{AB}$. It disappears if, and only if, $\rho_{AB}$ is separable for the $2\otimes2$ and $2\otimes3$ systems~\cite{Horodecki98}. For the purposes of this discussion, we use the following definition of negativity:
\begin{equation}
\mathcal{N}(\rho_{AB})=\|\rho^{T_A}_{AB}\|-1.
\end{equation}
For any maximally entangled state in a two-qubit system, this negativity is equal to 1.
CREN gives a perfect discrimination of positive partial transposition-bound entangled states and separable states in any bipartite quantum system~\cite{Horodeki97,Dur00}.
For any mixed state $\rho_{AB}$, CREN is defined as
\begin{equation}\label{eq:Corollary 5}
  \mathcal{N}_c(\rho_{AB})=\min\sum_{i}p_{i}\mathcal{N}(|\psi_i\rangle_{AB}),
\end{equation}
where the minimum is taken over all possible pure state
decompositions $\{p_{i},|\psi_i\rangle_{AB}\}$ of $\rho_{AB}$.

For any mixed state $\rho_{AB}$, CRENOA is defined as~\cite{Kim09}
\begin{equation}\label{eq:Corollary 6}
  \mathcal{N}_{a}(\rho_{AB})=\max\sum_{i}p_{i}\mathcal{N}(|\psi_i\rangle_{AB}),
\end{equation}
where the maximum is taken over all possible pure state
decompositions $\{p_{i},|\psi_i\rangle_{AB}\}$ of $\rho_{AB}$.

CREN is equivalent to concurrence for any pure state with Schmidt rank-2~\cite{Kim09},
and consequently, it follows that for any two-qubit mixed state $\rho_{AB}=\sum_ip_i|\psi_i\rangle\langle\psi_i|$:
\begin{equation}\label{eq:Corollary 7}
  \mathcal{N}_c(\rho_{AB})=\min\sum_{i}p_{i}\mathcal{N}(|\psi_i\rangle_{AB})=\min\sum_{i}p_{i}\mathcal{C}(|\psi_i\rangle_{AB})=\mathcal{C}(\rho_{AB})
\end{equation}
and
\begin{equation}\label{eq:Corollary 8}
  \mathcal{N}_{a}(\rho_{AB})=\max\sum_{i}p_{i}\mathcal{N}(|\psi_i\rangle_{AB})=\max\sum_{i}p_{i}\mathcal{C}(|\psi_i\rangle_{AB})=\mathcal{C}_{a}(\rho_{AB}),
\end{equation}
where the minimum and the maximum are taken over all pure state decompositions $\{p_{i},|\psi_i\rangle_{AB}\}$ of $\rho_{AB}$.

\section*{Monogamy relations of concurrence and negativity}\label{sec:MCN}

The CKW inequality~\cite{Coffman00} was first defined as
\begin{equation}\label{eq:Corollary 9}
  \mathcal{C}^2(\rho_{A|BC}) \geq \mathcal{C}^2(\rho_{AB})+\mathcal{C}^2(\rho_{AC}),
\end{equation}
where $\mathcal{C}(\rho_{A|BC}) $ is the concurrence of a three-qubit state $\rho_{A|BC}$ for any bipartite cut of subsystems between $A$ and $BC$.
Similarly, the dual inequality in terms of
COA is as follows~\cite{Gour05}:
\begin{equation}\label{eq:Corollary 10}
   \mathcal{C}^2(\rho_{A|BC})\leq\mathcal{C}_a^2(\rho_{AB})+\mathcal{C}_a^2(\rho_{AC}).
\end{equation}

For any pure state $|\psi\rangle_{A_{1}...A{n}}$ in an $n$-qubit system $A_1\otimes...\otimes A_n$, where $ A_i\cong C^2 $ for $i=1, ... ,n, $ a generalisation of the CKW inequality is
\begin{equation}\label{eq:Corollary 11}
  \mathcal{C}^2(|\psi\rangle_{A_{1}|A_{2}...A_{n}})\geq \mathcal{C}^2(\rho_{A_{1}A_{2}})+...+\mathcal{C}^2(\rho_{A_{1}A_{n}}).
\end{equation}
The dual inequality in terms of the
COA for $n$-qubit states has the form~\cite{Kim09}
\begin{equation}\label{eq:Corollary 12}
   \mathcal{C}^2(|\psi\rangle_{A_{1}|A_{2}...A_{n}}) \leq \mathcal{C}_a^2(\rho_{A_{1}A_{2}})+...+\mathcal{C}_a^2(\rho_{A_{1}A_{n}}).
\end{equation}

When the rank of the matrix is 2, we have
\begin{equation}\label{eq:Corollary 13}
   \mathcal{C}(|\psi\rangle_{A_{1}|A_{2}...A_{n}})= \mathcal{N}(|\psi\rangle_{A_{1}|A_{2}...A_{n}}).
\end{equation}
Combining Eq.~(\ref{eq:Corollary 7}) with Eq.~(\ref{eq:Corollary 8}), we have
\begin{equation}\label{eq:Corollary 14}
   \mathcal{C}(\rho_{A_{i}A_j})=\mathcal{N}_c(\rho_{A_{i}A_j}), ~~ \mathcal{C}_a(\rho_{A_{i}A_j})=\mathcal{N}_a(\rho_{A_{i}A_j}),
\end{equation}
where $i,j\in\{1,...,n\},i\neq j$.

For any $n$-qubit pure state $|\psi\rangle_{A_{1}...A{n}}$, we have
\begin{equation}\label{eq:Corollary 15}
  \mathcal{N}^2(|\psi\rangle_{A_{1}|A_{2}...A_{n}})\geq \mathcal{N}^2_c(\rho_{A_{1}A_{2}})+...+\mathcal{N}^2_c(\rho_{A_{1}A_{n}}).
\end{equation}
The dual inequality~\cite{Kim09} in terms of CRENOA is as follows:
\begin{equation}\label{eq:Corollary 16}
   \mathcal{N}^2(|\psi\rangle_{A_{1}|A_{2}...A_{n}}) \leq \mathcal{N}_a^2(\rho_{A_{1}A_{2}})+...+\mathcal{N}_a^2(\rho_{A_{1}A_{n}}).
\end{equation}

\section*{Monogamy inequalities of CREN}\label{sec:MN}

For a $2\otimes2\otimes m$ quantum pure state $|\psi\rangle_{ABC}$, it has been
shown that $\mathcal{C}_a^2(\rho_{AB})=\mathcal{C}^2(\rho_{AB})+\tau_2^C(|\psi\rangle_{ABC})$~\cite{Gour05}, where $\tau_2^C(|\psi\rangle_{ABC})=\mathcal{C}^2(|\psi\rangle_{A|BC})-\mathcal{C}^2(\rho_{AB})-\mathcal{C}^2(\rho_{AC})$ is the three-tangle of concurrence. $\mathcal{C}(|\psi\rangle_{A|BC})$ is the concurrence under bipartition
$A|BC$ for pure state $|\psi\rangle_{ABC}$. Namely,
\begin{equation}\label{eq:Corollary 17}
\mathcal{C}^2(|\psi\rangle_{A|BC_1...C_{n-2}})=\mathcal{C}_a^2(\rho_{AB})+\mathcal{C}^2(\rho_{A|C_1...C_{n-2}}).
\end{equation}
Similarly, considering that CREN is equivalent to concurrence by Eq.~(\ref{eq:Corollary 14}), we have
\begin{equation}\label{eq:Corollary 18}
\mathcal{N}^2(|\psi\rangle_{A|BC_1...C_{n-2}})=\mathcal{N}_a^2(\rho_{AB})+\mathcal{N}^2_c(\rho_{A|C_1...C_{n-2}}).
\end{equation}

The concurrence is related to the linear entropy of a
state~\cite{Santos00}
\begin{equation}\label{eq:Corollary 19}
  T(\rho )=1-\rm{Tr}(\rho^2).
\end{equation}

Given a bipartite state $\rho$ , $ T(\rho)$ has the property~\cite{Zhang08},
\begin{equation}\label{eq:Corollary 20}
  T(\rho_A )+  T(\rho_B )\geq T(\rho_{AB} )\geq|T(\rho_A )-T(\rho_B )|.
\end{equation}

From the definition of pure state concurrence in Eq.~(\ref{eq:Corollary 1})
together with Eq.~(\ref{eq:Corollary 19}), we have
\begin{equation}\label{eq:Corollary 24}
\mathcal{C}^2(|\psi_i\rangle_{AB|C_1...C_{n-2}})=2[1-\rm{Tr}(\rho_{AB}^2)]=2T(\rho_{AB}).
\end{equation}

Now, we provide the following theorems:

{\sf Theorem 1}~ For any $2\otimes 2\otimes 2$ tripartite mixed state $\rho_{ABC}$ we have
\begin{equation}\label{eq:Corollary 21}
 \mathcal{N}_a^2(\rho_{A|BC})\leq\mathcal{N}_a^2(\rho_{B|AC})+\mathcal{N}_a^2(\rho_{C|AB}).
\end{equation}

\emph{Proof.} Let $\rho_{ABC}=\sum_{i}p_{i}|\psi_i\rangle_{ABC}\langle\psi_i|$ be an optimal decomposition realising $\mathcal{N}_{a}(\rho_{A|BC})$; that is,
\begin{equation}\label{eq:Corollary 22}
  \mathcal{N}_{a}(\rho_{A|BC})=\max\sum_{i}p_{i}\mathcal{N}(|\psi_i\rangle_{A|BC}),
\end{equation}
where $\rho_{BC}=\rm{Tr}_A|\psi_i\rangle_{ABC}\langle\psi_i|$, $\rho_{B}=\rm{Tr}_{AC}|\psi_i\rangle_{ABC}\langle\psi_i|$ and  $\rho_{C}=\rm{Tr}_{AB}|\psi_i\rangle_{ABC}\langle\psi_i|$, and we have
\begin{equation}
  \mathcal{N}^2(|\psi_i\rangle_{A|BC})=\mathcal{C}^2(|\psi_i\rangle_{A|BC})=2T(\rho_{A})=2T(\rho_{BC}).
\end{equation}

Combining Eq.~(\ref{eq:Corollary 20}) with Eq.~(\ref{eq:Corollary 24}), we have
\begin{equation}\label{eq:Corollary 25}
\begin{aligned}
  2T(\rho_{BC})&\leq2T(\rho_B )+2T(\rho_C )\\
  &=\mathcal{C}^2(|\psi_i\rangle_{B|AC})+\mathcal{C}^2(|\psi_i\rangle_{C|AB})\\
   &=\mathcal{N}^2(|\psi_i\rangle_{B|AC}) + \mathcal{N}^2(|\psi_i\rangle_{C|AB}).
\end{aligned}
\end{equation}
The third equality holds because CREN and concurrence are equal for any rank-2 pure state. Therefore, we obtain
\begin{equation}\label{eq:Corollary 26}
  \mathcal{N}^2(|\psi_i\rangle_{A|BC})\leq\mathcal{N}^2(|\psi_i\rangle_{B|AC}) + \mathcal{N}^2(|\psi_i\rangle_{C|AB}).
\end{equation}

Combining Eq.~(\ref{eq:Corollary 22}) with Eq.~(\ref{eq:Corollary 26}), we finally get
\begin{equation}\label{eq:Corollary 27}
 \mathcal{N}_a^2(\rho_{A|BC})\leq\mathcal{N}_a^2(\rho_{B|AC})+\mathcal{N}_a^2(\rho_{C|AB}).
\end{equation}

Thus, the proof is completed.\qquad \qquad \qquad \qquad \qquad \qquad \qquad \qquad \qquad \qquad \qquad \qquad \qquad \qquad \qquad \qquad \qquad \qquad $\square$

Theorem~1 shows a simple relationship of CRENOA in a tripartite quantum system. The monogamy inequality shows that the entanglement $A|BC$ cannot be
greater than the sum of the entanglement $B|AC$ and the entanglement $C|AB$. Taking an easy example, when considering a three-qubit state, the following equation exists: $|\psi\rangle_{ABC}=a|010\rangle+b|100\rangle$ where $|a|^2+|b|^2=1$. Using a simple calculation, the following equation can be obtained: $\mathcal{N}_a^2(\rho_{A|BC})=\mathcal{N}_a^2(\rho_{B|AC})+\mathcal{N}_a^2(\rho_{C|AB})$ where the state $|\psi\rangle_{ABC}$ saturates the monogamy inequality in Eq.~(\ref{eq:Corollary 21}). 
Moreover, the iteration of Eq.~(\ref{eq:Corollary 21}) leads us to the generalized monogamy inequality in multi-qubit quantum systems.

{\sf Corollary 1}~ For any multi-party mixed state $\rho_{A_{1}|A_{2}...A_{n}}$ in an $n$-qubit system~\cite{Luo15}, the following monogamy inequality exists:
\begin{equation}
   \mathcal{N}_a^2(\rho_{A_{1}|A_{2}...A_{n}}) \leq \sum^n_{i=2}\mathcal{N}_a^2(\rho_{A_{i}|A_{1}...A_{i-1}A_{i+1}...A_{n}})\leq\sum_{i=2}^{n}\sum_{j=1,j\neq i}^{n}\mathcal{N}_a^2(\rho_{A_{i}A_{j}}).
\end{equation}

The meaning of the first inequality is clear the bipartite entanglement between $\rho_{A_1}$ and the other qubits, when taken as a group cannot be greater than the sum of the $n-1$ individual bipartite entanglements between $\rho_{A_i}~(i\neq1)$ and
the other remaining qubits.
We now start to consider a four-qubit system. As shown in Fig.~(a), the squared CRENOA with respect to the bipartition ($A|BCD$) is not greater than the sum of the three squared CRENOAs (the three possible bipartitions are $B|ACD$, $C|ABD$ and $D|ABC$).
\begin{figure}[ht]
\centering
\includegraphics[width=5in]{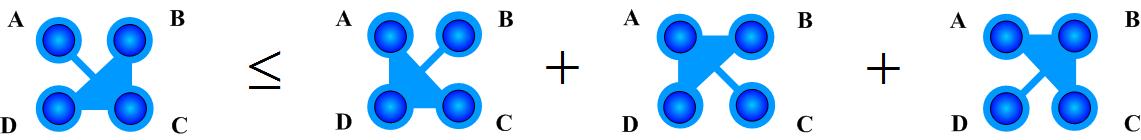}
\caption*{(a) The example shows the reciprocal relation of squared CRENOA in a four-qubit system.
}
\label{}
\end{figure}

The meaning of the second inequality is clear the sum of the bipartite entanglements between $\rho_{A_i}~(i\neq1)$ and
the other remaining qubits cannot be greater than the sum of the bipartite entanglements $\rho_{A_iA_j}~(i\neq1,j\neq i)$.

{\sf Theorem 2}~ For any $n$-qubit pure state $|\psi\rangle_{ABC_1...C_{n-2}}$, we have
\begin{equation}\label{eq:Corollary 28}
2\mathcal{N}_a^2(\rho_{AB})+\sum_{i=1}^{n-2}\mathcal{N}_a^2(\rho_{AC_i})+\sum_{i=1}^{n-2}\mathcal{N}_a^2(\rho_{BC_i})\geq  \mathcal{N}^2(|\psi\rangle_{AB|C_1...C_{n-2}}),
\end{equation}
where $\rho_{AB}=\rm{Tr}_{C_1...C_{n-2}}(|\psi\rangle\langle\psi|)$, $\rho_{AC_i}=\rm{Tr}_{BC_1...C_{i-1}C_{i+1}...C_{n-2}}(|\psi\rangle\langle\psi|)$ and $\rho_{BC_i}=\rm{Tr}_{AC_1...C_{i-1}C_{i+1}...C_{n-2}}(|\psi\rangle\langle\psi|)$.

\emph{Proof.} From  the result of Theorem~1, we find that the generalised monogamy inequality can be easily obtained by using the superposition of states. We now consider $\mathcal{N}^2(|\psi\rangle_{AB|C_1...C_{n-2}})$. When the rank of the matrix is 2, we have
\begin{equation}
  \mathcal{N}^2(|\psi\rangle_{AB|C_1...C_{n-2}})=\mathcal{C}^2(|\psi\rangle_{AB|C_1...C_{n-2}})=2T(\rho_{AB}).
\end{equation}

Combining Eq.~(\ref{eq:Corollary 20}) with Eq.~(\ref{eq:Corollary 24}), we get the relationship
\begin{equation}
 \begin{aligned}
2T(\rho_{AB})&\leq2T(\rho_A )+2T(\rho_B )\\
  &=\mathcal{C}^2(|\psi_i\rangle_{A|BC_1...C_{n-2}})+\mathcal{C}^2(|\psi_i\rangle_{B|AC_1...C_{n-2}})\\
   &=\mathcal{N}^2(|\psi_i\rangle_{A|BC_1...C_{n-2}})+\mathcal{N}^2(|\psi_i\rangle_{B|AC_1...C_{n-2}}).
\end{aligned}
\end{equation}
The third equality follows from the fact that CREN and concurrence are equal for any rank-2 pure state.
\begin{equation}\label{eq:Corollary 43}
  \mathcal{N}^2(|\psi_i\rangle_{AB|C_1...C_{n-2}})\leq\mathcal{N}^2(|\psi_i\rangle_{A|BC_1...C_{n-2}}) + \mathcal{N}^2(|\psi_i\rangle_{B|AC_1...C_{n-2}}).
\end{equation}
For a mixed state, CRENOA is expressed as $\mathcal{N}(|\psi_i\rangle_{A|BC_1...C_{n-2}})$, and we have
\begin{equation}
  \mathcal{N}_{a}(\rho_{A|BC_1...C_{n-2}})=\max\sum_{i}p_{i}\mathcal{N}(|\psi_i\rangle_{A|BC_1...C_{n-2}}).
\end{equation}

Furthermore, when combining this with Eq.~(\ref{eq:Corollary 43}), we finally get
\begin{equation}\label{eq:Corollary 29}
 \mathcal{N}^2(|\psi\rangle_{AB|C_1...C_{n-2}})\leq\mathcal{N}_a^2(\rho_{A|BC_1...C_{n-2}})+\mathcal{N}_a^2(\rho_{B|AC_1...C_{n-2}})
\end{equation}
and
\begin{equation}\label{eq:Corollary 30}
\begin{aligned}
\mathcal{N}_a^2(\rho_{A|BC_1...C_{n-2}})\leq\mathcal{N}_a^2(\rho_{AB})+\sum_{i=1}^{n-2}\mathcal{N}_a^2(\rho_{AC_i}),\\
\mathcal{N}_a^2(\rho_{B|AC_1...C_{n-2}})\leq\mathcal{N}_a^2(\rho_{BA})+\sum_{i=1}^{n-2}\mathcal{N}_a^2(\rho_{BC_i}).
\end{aligned}
\end{equation}
Combining Eq.~(\ref{eq:Corollary 29}) with Eq.~(\ref{eq:Corollary 30}), we have Eq.~(\ref{eq:Corollary 28}). In other words, we give an upper bound about $\mathcal{N}^2(|\psi\rangle_{AB|C_1...C_{n-2}})$, i.e.,
\begin{equation}\label{eq:Corollary 31}
  2\mathcal{N}_a^2(\rho_{AB})+\sum_{i=1}^{n-2}\mathcal{N}_a^2(\rho_{AC_i})+\sum_{i=1}^{n-2}\mathcal{N}_a^2(\rho_{BC_i})\geq  \mathcal{N}^2(|\psi\rangle_{AB|C_1...C_{n-2}}).
\end{equation}
This completes the proof. \qquad \qquad \qquad \qquad \qquad \qquad \qquad \qquad \qquad \qquad \qquad \qquad \qquad \qquad \qquad \qquad \qquad \qquad \qquad $\square$

Theorem~2 shows that the entanglement between $AB$ and the other qubits cannot be
greater than the sum of the individual entanglements between $A$ and each of the $n-1$ remaining qubits and the individual
entanglements between $B$ and each of the $n-1$ remaining qubits. Theorem~2 provides a polygamy-type upper bound of multi-qubit entanglement between the two-qubit system $AB$ and the other $(n-2)$-qubit system $C_1C_2...C_{n-2}$ in terms of the squared CRENOA. Especially under partition $AB$,
a two-qubit system is different from the previous monogamy inequality.
When $|\psi\rangle_{AB|C_1...C_{n-2}}=|\psi\rangle_{A}\otimes|\psi\rangle_{B|C_1...C_{n-2}}$,
the calculation results in $\mathcal{N}_a^2(\rho_{AB})=0,\mathcal{N}_a^2(\rho_{AC_i})=0$. Consequently, the polygamy-type relation is obtained as shown in Eq.~(\ref{eq:Corollary 16}).

Finally, consider the following four-qubit state: $|\psi\rangle_{ABCD}=a|0100\rangle+b|0010\rangle+c|0001\rangle$ where $|a|^2+|b|^2+|c|^2=1$. We can easily get the following equations: $\mathcal{N}_a^2(\rho_{AB})=\mathcal{N}_a^2(\rho_{AC})=\mathcal{N}_a^2(\rho_{AD})=0$ and $\mathcal{N}_a^2(\rho_{BC})+\mathcal{N}_a^2(\rho_{BD})= \mathcal{N}^2(|\psi\rangle_{AB|CD})=\frac{16}{9}$. Therefore, the state $|\psi\rangle_{ABCD}$
saturates the monogamy inequality in Eq.~(\ref{eq:Corollary 28}).

{\sf Theorem 3}~ For any $n$-qubit pure state $|\psi\rangle_{ABC_1...C_{n-2}}$,
\begin{equation}\label{eq:Corollary 44}
\mathcal{N}^2(|\psi\rangle_{AB|C_1...C_{n-2}})\geq |\sum_{i=1}^{n-2}\mathcal{N}_a^2(\rho_{AC_i})-\sum_{i=1}^{n-2}\mathcal{N}_a^2(\rho_{BC_i})|,
\end{equation}
where $\rho_{AB}=\rm{Tr}_{C_1...C_{n-2}}(|\psi\rangle\langle\psi|)$, $\rho_{AC_i}=\rm{Tr}_{BC_1...C_{i-1}C_{i+1}...C_{n-2}}(|\psi\rangle\langle\psi|)$ and  $\rho_{BC_i}=\rm{Tr}_{AC_1...C_{i-1}C_{i+1}...C_{n-2}}(|\psi\rangle\langle\psi|)$.

\emph{Proof.} We have the following property for linear entropy~\cite{Zhang08}:
\begin{equation}\label{eq:Corollary 32}
T(\rho_{AB} )\geq|T(\rho_A )-T(\rho_B )|.
\end{equation}

Combining Eq.~(\ref{eq:Corollary 24}) with Eq.~(\ref{eq:Corollary 32}), we have
\begin{equation}
2[1-\rm{Tr}(\rho_{AB}^2)]\geq|2[1-\rm{Tr}(\rho_{A}^2)]-2[1-\rm{Tr}(\rho_{B}^2)]|
\end{equation}
and
\begin{equation}
\mathcal{C}^2(|\psi\rangle_{AB|C_1...C_{n-2}})\geq |\mathcal{C}^2(|\psi\rangle_{A|BC_1...C_{n-2}})-\mathcal{C}^2(|\psi\rangle_{B|AC_1...C_{n-2}})|.
\end{equation}

By using the equivalent relation between concurrence and CREN (see Eq.~(\ref{eq:Corollary 14})), we have
\begin{equation}\label{eq:Corollary 51}
\mathcal{N}^2(|\psi\rangle_{AB|C_1...C_{n-2}})\geq |\mathcal{N}^2(|\psi\rangle_{A|BC_1...C_{n-2}})-\mathcal{N}^2(|\psi\rangle_{B|AC_1...C_{n-2}})|.
\end{equation}

There is a relationship between CREN and CRENOA (see Eq.~(\ref{eq:Corollary 18})):
\begin{equation}
   \mathcal{N}^2(|\psi\rangle_{A|BC_1...C_{n-2}}) = \mathcal{N}_a^2(\rho_{AB})+\mathcal{N}^2_c(\rho_{A|C_1...C_{n-2}})
\end{equation}
\begin{equation}
   \mathcal{N}^2(|\psi\rangle_{B|AC_1...C_{n-2}}) = \mathcal{N}_a^2(\rho_{BA})+\mathcal{N}^2_c(\rho_{B|C_1...C_{n-2}}).
\end{equation}
Putting the above two equalities into Eq.~(\ref{eq:Corollary 51}), we get
\begin{eqnarray}
|\mathcal{N}^2_c(\rho_{A|C_1...C_{n-2}})-\mathcal{N}^2_c(\rho_{B|C_1...C_{n-2}})| \geq|\sum_{i=1}^{n-2}\mathcal{N}_c^2(\rho_{AC_i})-\sum_{i=1}^{n-2}\mathcal{N}_a^2(\rho_{BC_i})|\geq|\sum_{i=1}^{n-2}\mathcal{N}_a^2(\rho_{AC_i})-\sum_{i=1}^{n-2}\mathcal{N}_a^2(\rho_{BC_i})|.
\end{eqnarray}

Similar to the above derivation, we give a lower bound about $\mathcal{N}^2(|\psi\rangle_{AB|C_1...C_{n-2}})$, i.e.,
\begin{equation}
\mathcal{N}^2(|\psi\rangle_{AB|C_1...C_{n-2}})\geq |\sum_{i=1}^{n-2}\mathcal{N}_a^2(\rho_{AC_i})-\sum_{i=1}^{n-2}\mathcal{N}_a^2(\rho_{BC_i})|.
\end{equation}

This lower bound is a direct consequence of CREN. \qquad \qquad \qquad \qquad \qquad \qquad \qquad \qquad \qquad \qquad \qquad \qquad \qquad $\square$

Theorem~3 shows that the entanglement between $AB$ and the other qubits cannot be
less than the absolute value of the difference between both the individual entanglements between $A$ and each of the $n-1$ remaining qubits and the individual
entanglements between $B$ and each of the $n-1$ remaining qubits.
Theorem~3 provides a monogamy-type lower bound of multi-qubit entanglement between the two-qubit system $AB$ and the other $(n-2)$-qubit system $C_1C_2...C_{n-2}$ in terms of the squared CRENOA.
When $|\psi\rangle_{AB|C_1...C_{n-2}}=|\psi\rangle_{B}\otimes|\psi\rangle_{A|C_1...C_{n-2}}$, $\mathcal{N}_a^2(\rho_{BC_i})=0$, and so we obtain the CWK-type relation in Eq.~(\ref{eq:Corollary 15}).

Finally, we consider the following four-qubit state
$|\psi\rangle_{ABCD}=a|1000\rangle+b|0010\rangle+c|0001\rangle$ where $|a|^2+|b|^2+|c|^2=1$, from which we can easily obtain the following equations: $\mathcal{N}_a^2(\rho_{BC})=\mathcal{N}_a^2(\rho_{BD})=0$ and $\mathcal{N}_a^2(\rho_{AC})+\mathcal{N}_a^2(\rho_{AD})= \mathcal{N}^2(|\psi\rangle_{AB|CD})=\frac{16}{9}$. Therefore, the state $|\psi\rangle_{ABCD}$
saturates the monogamy inequality in Eq.~(\ref{eq:Corollary 44}). Therefore, a generalised monogamy inequality using negativity and CRENOA in an $n$-qubit is proposed. These relations also give rise to a type of trade-off in inequalities that is related to the upper and lower bounds of CRENOA.

{\sf Remark }~ It is interesting to note that the properties of CREN are based on the subadditivity of linear entropy. However, negativity violates this subadditivity in general conditions~\cite{Rossignoli10,Hu06,Rastegin11}.

\section*{Examples}\label{sec:W-C}
In this section, we use some special states to study generalised monogamy inequalities. 
First, we consider the (Greenberger-Horne-Zeilinger) GHZ state and W state in Examples 1 and 2. Second, we consider two states in the higher-dimensional system in Examples 3 and 4.

Example 1. For an arbitrary pure GHZ state in an $n$-qubit system:
\begin{equation}
  |GHZ\rangle=a|0\rangle^{\otimes n}+b|1\rangle^{\otimes n},
\end{equation}
where $|a|^2+|b|^2=1$. The generalized GHZ state is satisfied with the previous CKW inequality. We will now show that the generalised GHZ state satisfies the generalised monogamy inequalities. We have $ \rho_1=\rho_2=...=\rho_n=a^2|0\rangle\langle0|+b^2|1\rangle\langle1|$. It is straightforward to check: $\mathcal{N}_a^2(\rho_{A_{1}|A_{2}...A_{n}})=\mathcal{N}_a^2(\rho_{A_{2}|A_{1}...A_{n}})=...=\mathcal{N}_a^2(\rho_{A_{i}|A_{1}...A_{i-1}A_{i+1}...A_{n}}) = 4|(ab)^2|$ and
$\mathcal{N}^2_a(\rho_{A_{1}A_{2}})=\mathcal{N}^2_a(\rho_{A_{2}A_{3}})=...=\mathcal{N}^2_a(\rho_{A_{i}A_{i+1}})= 4|ab|$,
$\mathcal{N}^2(\rho_{A_{1}A_{2}|A_{3}...A_{n}})= 4|(ab)^2|$. Therefore:
\begin{equation}
   \mathcal{N}_a^2(\rho_{A_{1}|A_{2}...A_{n}}) = 4|(ab)^2| \leq 4(n-1)|(ab)^2| = \sum^n_{i=2}\mathcal{N}_a^2(\rho_{A_{i}|A_{1}...A_{i-1}A_{i+1}...A_{n}}),
\end{equation}
\begin{equation}
2\mathcal{N}_a^2(\rho_{AB})+\sum_{i=1}^{n-2}\mathcal{N}_a^2(\rho_{AC_i})+\sum_{i=1}^{n-2}\mathcal{N}_a^2(\rho_{BC_i}) = 8(n-1)|(ab)^2| \geq 4|(ab)^2| = \mathcal{N}^2(|\psi\rangle_{AB|C_1...C_{n-2}}),
\end{equation}
\begin{equation} 
\mathcal{N}^2(|\psi\rangle_{AB|C_1...C_{n-2}}) = 4|(ab)^2| \geq 0 = |\sum_{i=1}^{n-2}\mathcal{N}_a^2(\rho_{AC_i})-\sum_{i=1}^{n-2}\mathcal{N}_a^2(\rho_{BC_i})|.
\end{equation}

Example 2. For a pure state $|W\rangle$ in an $n$-qubit system:
\begin{equation}
  |W\rangle=\frac{1}{\sqrt{n}}(|10...0\rangle+|01...0\rangle...+|00...1\rangle),
\end{equation}
with $\sum_{i=1}^n|\frac{1}{\sqrt{n}}|^2=1$. It is very important to understand the saturation of the previous CKW inequality. Using a simple calculation, we have $\rho_1=\rho_2=...=\rho_n=\frac{1}{n}(|1\rangle\langle1|)+\frac{n-1}{n}(|0\rangle\langle0|)$. It is straightforward to check: $\mathcal{N}_a^2(\rho_{A_{1}|A_{2}...A_{n}})=\mathcal{N}_a^2(\rho_{A_{2}|A_{1}...A_{n}}) =...= \mathcal{N}_a^2(\rho_{A_{i}|A_{1}...A_{i-1}A_{i+1}...A_{n}}) = \frac{4(n-1)}{n^2}.$
$\mathcal{N}^2_a(\rho_{A_{1}A_{2}}) = \mathcal{N}^2_a(\rho_{A_{2}A_{3}}) =...= \mathcal{N}^2_a(\rho_{A_{i}A_{i+1}}) = \frac{4}{n^2}$,
$\mathcal{N}^2(\rho_{A_{1}A_{2}|A_{3}...A_{n}})= \frac{8(n-2)}{n^2}$. In the same way, we get the following inequalities:
\begin{equation}
   \mathcal{N}_a^2(\rho_{A_{1}|A_{2}...A_{n}}) = \frac{4(n-1)}{n^2} \leq \frac{4(n-1)^2}{n^2} = \sum^n_{i=2}\mathcal{N}_a^2(\rho_{A_{i}|A_{1}...A_{i-1}A_{i+1}...A_{n}}),
\end{equation}
\begin{equation}
2\mathcal{N}_a^2(\rho_{AB})+\sum_{i=1}^{n-2}\mathcal{N}_a^2(\rho_{AC_i})+\sum_{i=1}^{n-2}\mathcal{N}_a^2(\rho_{BC_i}) = \frac{8(n-1)}{n^2} \geq \frac{8(n-2)}{n^2} = \mathcal{N}^2(|\psi\rangle_{AB|C_1...C_{n-2}}),
\end{equation}
\begin{equation}
\mathcal{N}^2(|\psi\rangle_{AB|C_1...C_{n-2}}) = \frac{8(n-2)}{n^2} \geq 0 = |\sum_{i=1}^{n-2}\mathcal{N}_a^2(\rho_{AC_i})-\sum_{i=1}^{n-2}\mathcal{N}_a^2(\rho_{BC_i})|.
\end{equation}

From the above results, we discover that the generalised GHZ state and W state satisfy our inequalities. We further explore the condition of the generalised inequalities in higher-dimensional systems. We consider the following examples:

Example 3. For a pure, totally antisymmetric state $|\psi_{ABC}\rangle$ in a $3\otimes3\otimes3 $ system~\cite{Ou07}:
\begin{equation}
  |\psi_{ABC}\rangle=\frac{1}{\sqrt{6}}(|123\rangle-|132\rangle+|231\rangle-|213\rangle+|312\rangle-|321\rangle).
  \end{equation}

This special quantum state is not satisfied with the previous CKW inequality~\cite{ch00} but it is established in generalised monogamy inequalities. We can easily obtain
$\mathcal{N}^2_a(\rho_{A|BC})=\mathcal{N}^2_a(\rho_{B|AC})=\mathcal{N}^2_a(\rho_{C|AB})=4$ and further obtain the inequalities $\mathcal{N}_a^2(\rho_{A|BC})\leq\mathcal{N}_a^2(\rho_{B|AC})+\mathcal{N}_a^2(\rho_{C|AB})$.
We now explore theorems~2 and~3. First, we have $\mathcal{N}^2_a(\rho_{AB})=1,\mathcal{N}^2_a(\rho_{AC})=1,\mathcal{N}^2_a(\rho_{BC})=1$ and $\mathcal{N}^2(|\psi\rangle_{AB|C})=4 $. Therefore, we obtain the following inequalities:
\begin{equation}    
2\mathcal{N}^2_a(\rho_{AB})+\mathcal{N}^2_a(\rho_{AC})+\mathcal{N}^2_a(\rho_{BC})\geq\mathcal{N}^2(|\psi\rangle_{AB|C})\geq | \mathcal{N}^2_a(\rho_{AC})- \mathcal{N}^2_a(\rho_{BC})|.
\end{equation}

Example 4. The $n$-qudit generalised W-class state in higher-dimensional quantum systems is very useful in quantum information theory~\cite{Lvzhou07}. We verify whether the generalised monogamy inequalities hold in higher-dimensional systems using a special example.
First, we recall the definition of $n$-qudit generalised W-class state~\cite{Kim08},
\begin{equation}
|W_n^d\rangle_{A_1...A_n}=\sum^{d-1}_{i=1}(a_{1i}|i0...0\rangle+a_{2i}|0i...0\rangle+...+a_{ni}|00...i\rangle),
\end{equation}
where $\sum^{n}_{s=1}\sum^{d-1}_{i=1}|a_{si}|^2=1$.

Let $|\psi\rangle_{A_1...A_n} $ be an $n$-qudit pure state in a
superposition of an $n$-qudit generalised W-class state and vacuum; that is,
\begin{equation}
|\psi\rangle_{A_1...A_n}=\sqrt{p}|W_n^d\rangle_{A_1...A_n}+\sqrt{1-p}|0...0\rangle_{A_1...A_n},
\end{equation}
for some $0\leq p \leq 1$.

For the squared negativity $\mathcal{N}^2 $ of $|\psi\rangle_{A_1...A_n}$ with respect to the bipartition between $A_1$ and the other qudits, the reduced
density matrix $\rho_{A_1}$ of $|\psi\rangle_{A_1...A_n}$ onto subsystem $A_1$ is obtained
as
\begin{eqnarray}\label{eq:Corollary 34}
  \rho_{A_1} &=&\rm{Tr}_{A_2...A_n}|\psi\rangle_{A_1A_2...A_n}\langle\psi|\nonumber \\
   &=& p\sum^{d-1}_{i,j=1}a_{1i}a_{1j}^*|i\rangle_{A_1}\langle j|
+\big[p\Omega+(1-p)\big]|0\rangle_{A_1}\langle0|+\sqrt{p(1-p)}\big[\sum^{d-1}_{i=1}a_{1i}|i\rangle_{A_1}\langle 0|+\sum^{d-1}_{j=1}a_{1j}^*|0\rangle_{A_1}\langle j|\big],
\end{eqnarray}
where $\Omega=\sum_{s=2}^{n}\sum^{d-1}_{i=1}|a_{si}|^2=1-\sum^{d-1}_{j=1}|a_{1j}|^2$.

When considering the $|\psi\rangle_{A_1|A_2...A_n}$ state, we need to obtain the eigenvalue of the matrix by applying the definition of pure state negativity in Eq.~(\ref{eq:Corollary 4}). Using a simple calculation, we find that the matrix has rank-2 and we have
\begin{equation}\label{eq:Corollary 36}
\mathcal{N}^2(|\psi\rangle_{A_1|A_2...A_n})=[(\rm{Tr}\sqrt{\rho_{A_1}})^2-1]^2=4\lambda_i\lambda_j=4p^2(1-\Omega)\Omega.
\end{equation}

We now consider the case in which $ n = 2 $. The remaining cases follow analogously. The two-qudit reduced density matrix $\rho_{A_1A_2}$ of $|\psi\rangle_{A_1A_2...A_n}$  is
obtained as
\begin{equation}\label{eq:Corollary 37}
\begin{aligned}
\rho_{A_1A_2}=&\rm{Tr}_{A_3...A_n}|\psi\rangle_{A_1A_2...A_n}\langle\psi|\\
=&p\sum^{d-1}_{i,j=1}\big[a_{1i}a_{1j}^*|i0\rangle_{A_1A_2}\langle j0|+a_{1i}a_{2j}^*|i0\rangle_{A_1A_2}\langle 0j|+a_{2i}a_{1j}^*|0i\rangle_{A_1A_2}\langle j0|+a_{2i}a_{2j}^*|0i\rangle_{A_1A_2}\langle 0j|\big]\\ &+(p\Omega_2+1-p)|00\rangle_{A_1A_2}\langle 00|+ \sqrt{p(1-p)}\sum^{d-1}_{k=1}\big[(a_{1k}|k0\rangle+a_{2k}|0k\rangle)_{A_1A_2}\langle 00|+|00\rangle_{A_1A_2}(a_{1k}^*\langle k0|+a_{2k}^*\langle 0k|)\big],
\end{aligned}
\end{equation}
where $\Omega_2=1-\sum^{d-1}_{j=1}(a_{1j}^2+a_{2j}^2)$.
For convenient calculation, we consider two unnormalised states:
\begin{equation}
|\widetilde{x}\rangle=\sqrt{p}\sum_{i=1}^{d-1}(a_{1i}|i0\rangle_{A_1A_2}+a_{2i}|0i\rangle_{A_1A_2})+\sqrt{1-p}|00\rangle_{A_1A_2},  |\widetilde{y}\rangle=\sqrt{\Omega_2}|00\rangle_{A_1A_2}.
\end{equation}

Consequently, $\rho_{A_1A_2}$ can be represented as $ \rho_{A_1A_2}=|\widetilde{x}\rangle_{A_1A_2}\langle\widetilde{x}|+|\widetilde{y}\rangle_{A_1A_2}\langle\widetilde{y}|,$
where $ |\widetilde{x}\rangle $ and $ |\widetilde{y}\rangle $ are unnormalised states of the subsystems $A_1A_2$. By the HJW theorem~\cite{Hughston93}, any pure-state decomposition
$\rho_{A_1A_2}=\sum_{h}^r|\widetilde{\psi}_{h}\rangle_{A_1A_2}\langle\widetilde{\psi}_{h}|$, with size $r>2$ can be obtained by an $r\times r$ unitary matrix $u_{hl}$ such that
\begin{equation}
  |\widetilde{\psi}_{h}\rangle_{A_1A_2}=u_{hl}|\widetilde{x}\rangle_{A_1A_2}+u_{h2}|\widetilde{y}\rangle_{A_1A_2}
\end{equation}
for each $h$, for the normalized state $|\psi_{h}\rangle_{A_1A_2}=|\widetilde{\psi}_{h}\rangle_{A_1A_2}/\sqrt{p_h}$ with $p_h=|\langle\widetilde{\psi}_{h}|\widetilde{\psi}_{h}\rangle|$.

We apply the definition of mixed state negativity in Eq.~(\ref{eq:Corollary 5}) and Eq.~(\ref{eq:Corollary 37}), and then we have the two-tangle based on the CREN of $ \rho_{A_1A_2} $ as
\begin{equation}\label{eq:Corollary 41}
\mathcal{N}^2_c(\rho_{A_1A_2})=\min\sum_{i}p_{i}\mathcal{N}^2(|\psi_i\rangle_{A_1A_2})
=4p^2(1-\Omega)\sum^{d-1}_{i=1}|a_{2i}|^2=4p^2(1-\Omega)\Omega',
\end{equation}
where $\Omega'=\sum^{d-1}_{i=1}|a_{2i}|^2$.

From the definition of pure state negativity in Eq.~(\ref{eq:Corollary 6}) and Eq.~(\ref{eq:Corollary 37}), we have
\begin{equation}\label{eq:Corollary 38}
\mathcal{N}^2_c(\rho_{A_1A_2|A_3...A_n})=\min\sum_{i}p_{i}\mathcal{N}^2(|\psi_i\rangle_{A_1A_2|A_3...A_n})=4p\Omega_2(1-\Omega_2).
\end{equation}

We now try to verify the generalised monogamy inequalities of CREN in an $n$-qudit system.
For convenient calculation, we assume that $\sum^{d-1}_{i=1}a_{1i}^2=a$, $\sum^{d-1}_{i=1}a_{2i}^2=b$, $\sum^{d-1}_{i=1}a_{1i}^4=A$, $\sum^{d-1}_{i=1}a_{2i}^4=B.$

We first consider the generalisation of Theorem~1.
\begin{equation}
 \mathcal{N}_a^2(\rho_{A_{1}|A_{2}...A_{n}})=\mathcal{N}_a^2(\rho_{A_{2}|A_{1}...A_{n}})=...=\mathcal{N}_a^2(\rho_{A_{i}|A_{1}...A_{i-1}A_{i+1}...A_{n}})=4p^2(1-\Omega)\Omega=4p^2(1-a)a.
\end{equation}
This special quantum state is satisfied with the generalised monogamy inequality in Eq.~(\ref{eq:Corollary 21}) i.e.,
\begin{equation}
   \mathcal{N}_a^2(\rho_{A_{1}|A_{2}...A_{n}}) \leq \sum^n_{i=2}\mathcal{N}_a^2(\rho_{A_{i}|A_{1}...A_{i-1}A_{i+1}...A_{n}}).
\end{equation}

For the generalisation of Theorem~2, the left of Eq.~(\ref{eq:Corollary 28}) is
\begin{equation}
  2\mathcal{N}_a^2(\rho_{AB})+\sum_{i=1}^{n-2}\mathcal{N}_a^2(\rho_{AC_i})+\sum_{i=1}^{n-2}\mathcal{N}_a^2(\rho_{BC_i}).
\end{equation}

Using Eq.~(\ref{eq:Corollary 5}) and Eq.~(\ref{eq:Corollary 36}) we can simplify the calculation to
\begin{equation}
  \mathcal{N}_a^2(\rho_{AB})+\sum_{i=1}^{n-2}\mathcal{N}_a^2(\rho_{AC_i})=\mathcal{N}^2_c(\rho_{A|BC_{1}...C_{n}})=4p^2(1-\Omega)\Omega=4p^2(1-a)a
\end{equation}
and
\begin{equation}\label{eq:Corollary 39}
  \sum_{i=1}^{n-2}\mathcal{N}_a^2(\rho_{BC_i})=\mathcal{N}^2_c(\rho_{B|C_{1}...C_{n}})=4p^2(1-\Omega')\Omega'=4p^2(1-b)b.
\end{equation}
After some calculations, we have
\begin{eqnarray}
 &&2\mathcal{N}_a^2(\rho_{AB})+\sum_{i=1}^{n-2}\mathcal{N}_a^2(\rho_{AC_i})+\sum_{i=1}^{n-2}\mathcal{N}_a^2(\rho_{BC_i})\nonumber=\mathcal{N}_a^2(\rho_{AB})+\sum_{i=1}^{n-2}\mathcal{N}_a^2(\rho_{AC_i})+\sum_{i=1}^{n-2}\mathcal{N}_a^2(\rho_{BC_i})+\mathcal{N}_a^2(\rho_{AB})\nonumber\\
 &=&\mathcal{N}^2_c(\rho_{A|BC_{1}...C_{n}})+\mathcal{N}^2_c(\rho_{B|C_{1}...C_{n}})+\mathcal{N}_a^2(\rho_{AB})=4p^2(1-a)a+4p^2(1-b)b+4p^2ab.
\end{eqnarray}

Second, taking Eq.~(\ref{eq:Corollary 38}) to the right side of Eq.~(\ref{eq:Corollary 28}), we then have
\begin{eqnarray}\label{eq:Corollary 40}
 \mathcal{N}^2(|\psi\rangle_{AB|C_1...C_{n-2}})=4p\Omega_2(1-\Omega_2)=4p^2[1-(a+b)](a+b).
\end{eqnarray}

After a straightforward calculation, we obtain
\begin{equation}
   2\mathcal{N}_a^2(\rho_{AB})+\sum_{i=1}^{n-2}\mathcal{N}_a^2(\rho_{AC_i})+\sum_{i=1}^{n-2}\mathcal{N}_a^2(\rho_{BC_i})- \mathcal{N}^2(|\psi\rangle_{AB|C_1...C_{n-2}})=12p^2ab\geq0.
\end{equation}

Therefore, this $n$-qudit pure state is satisfied with the generalised monogamy inequality in Eq.~(\ref{eq:Corollary 28}). In other words, the test of the Theorem~2 has been accomplished.
Next, we verify Theorem~3. First, we consider the term CREN from Eq.~(\ref{eq:Corollary 44}):
\begin{equation}\label{eq:Corollary 42}
  \sum_{i=1}^{n-2}\mathcal{N}_a^2(\rho_{AC_i})=\mathcal{N}^2_c(\rho_{A|BC_{1}...C_{n}})- \mathcal{N}_a^2(\rho_{AB})=4p^2(1-\Omega)\Omega-4p^2(1-\Omega)\Omega'=4p^2a(1-a-b).
\end{equation}

Calculating the absolute value of the difference between Eq.~(\ref{eq:Corollary 39}) and Eq.~(\ref{eq:Corollary 42}), we obtain
\begin{eqnarray}
  &&|\sum_{i=1}^{n-2}\mathcal{N}_a^2(\rho_{AC_i})-\sum_{i=1}^{n-2}\mathcal{N}_a^2(\rho_{BC_i})|=|4p^2(a-a^2-ab+b^2-b)|=4p^2(a-a^2-ab+b^2-b).
\end{eqnarray}
It is easy to check $4p^2(a-a^2-ab+b^2-b)>0$, as
\begin{eqnarray}
  &0\leq a+b\leq1 \Rightarrow a(a+b)\leq a\Rightarrow a^2+ab-b+b^2<a(a+b)\leq a\Rightarrow a^2+ab-b+b^2-a<0\\ &\Rightarrow a-a^2-ab+b^2-b>0\Rightarrow4p^2(a-a^2-ab+b^2-b)>0.\nonumber
\end{eqnarray}

After a straightforward calculation, we have
\begin{eqnarray}
  &&\mathcal{N}^2(|\psi\rangle_{AB|C_1...C_{n-2}}) -|\sum_{i=1}^{n-2}\mathcal{N}_a^2(\rho_{AC_i})-\sum_{i=1}^{n-2}\mathcal{N}_a^2(\rho_{BC_i})|=4p^2b(2-2b-a)\geq0.
\end{eqnarray}

Therefore, this $n$-qudit pure state satisfies the generalised monogamy inequality in Eq.~(\ref{eq:Corollary 44}). We have now verified the generalised monogamy inequalities. In other words, the generalised monogamy inequality are satisfied with the $n$-qudit pure state for all three of our theorems.

\section*{Conclusions}
In this paper, we have used CREN to study different types of monogamy relations. In particular, we have shown that CREN satisfies the generalised monogamy inequalities. We have investigated the CKW-like inequalities and generalised monogamy inequalities. Furthermore, the generalised monogamy inequalities related to CREN and CRENOA were obtained by $n$-qubit states. These relations also give rise to a type of trade-off in inequalities that is related to the upper and lower bounds of CRENOA. Finally, we have shown that the partially coherent superposition of the
generalised W-class state and vacuum extensions of CREN satisfies the generalised
monogamy inequalities. We believe that the generalised monogamy inequalities can be useful in quantum information theory.
This paper was based on the linear entropy. To continue this work, we will study the nature of other entropy further in the future work. We hope that our work will be useful to the quantum physics.

\section*{Acknowledgements}

It is a pleasure to thank F.G.Zhang for inspiring discussions. We thank the anonymous referees for their valuable comments. This work was supported by the National Nature Science Foundation of China (Grant No.1127123), the Higher School Doctoral Subject Foundation of Ministry of Education of China (Grant No. 20130202110001) and the Fundamental Research Funds for the Central Universitie (Grant No. 2016CBY003).

\section*{Author contributions statement}
T.T. and Y.Luo contributed the idea. T.T. performed the calculations and wrote the main manuscript. Y.Luo checked the calculations. Y.Li improved the manuscript. All authors contributed to the discussion and reviewed the manuscript.

\section*{Additional information}
Competing financial interests: The authors declare no competing financial interests.

\end{document}